\documentclass[11pt]{article}
\usepackage[dvips]{graphicx}  
\begin{document}

\title{\LARGE \bf  PROCESS PHYSICS:  From  QUANTUM FOAM to GENERAL
RELATIVITY }  
\author{{Reginald T. Cahill}\\
  {School of Chemistry, Physics and Earth Sciences}\\{ Flinders University
}\\ { GPO Box
2100, Adelaide 5001, Australia }\\{(Reg.Cahill@flinders.edu.au)}}

\date{}
\maketitle

\begin{abstract}
{\it Progress in the new information-theoretic process physics\footnote{{\bf Process Physics} Web
Page: http://www.socpes.flinders.edu.au/people/rcahill/processphysics.html} is reported in
which the link to the phenomenology of general relativity is made. In process physics the
fundamental assumption is that  reality is to be modelled as self-organising  semantic (or
internal or relational) information using a self-referentially limited neural network
model.  Previous progress in process physics included the demonstration that space and
quantum physics are emergent  and unified, with time  a distinct non-geometric process,
that   quantum phenomena are caused by fractal topological   defects  embedded in and
forming a growing three-dimensional fractal process-space, which is essentially a quantum
foam.   Other   features of the emergent physics were:    quantum field theory with
emergent flavour and confined colour, limited causality and  the Born quantum measurement
metarule,  inertia, time-dilation effects, gravity and the equivalence principle, a growing
universe with  a cosmological constant,  black holes and event horizons, and the emergence
of classicality.  Here general relativity and the technical language of general covariance is seen not to be
fundamental but a phenomenological construct, arising as an amalgam of two distinct phenomena: the
`gravitational' characteristics of the emergent quantum foam for which `matter' acts as a
sink, and the classical `spacetime' measurement protocol, but with the later violated by
quantum measurement processes.  Quantum gravity, as manifested in the emergent Quantum Homotopic Field Theory of
the process-space or quantum foam,  is logically prior to the emergence  of the general relativity
phenomenology, and cannot be derived from it.  }

\end{abstract}

\vspace{20mm}
  Key words:  process physics, neural network, semantic information,
self-referential 

\hspace{17mm}  noise, process-time,  quantum foam, general relativity, quantum gravity.

\newpage

\section{Introduction}

 {\it Process Physics}    \cite{RC01,CK97,CK98,CK99,CKK00,MC} is a radical  information-theoretic 
modelling of reality  which  arose from analysis  of various extant limitations; from
the limitations of formal information systems discovered by G\"{o}del, Turing and
Chaitin,  from the limitations of the geometric modelling of time in the models constructed
by Galileo, Newton and Einstein, and by the limitations of the quantum theory and its
failure to account for the measurement process.  As is usual within physics these 
limitations were obscured by various metarules and metaphysical stories that have 
become treasured folklore, particularly the curved spacetime construct which is the
particular subject of this work.   

In process physics the
fundamental assumption is that  reality is to be modelled as self-organising  semantic 
information, that is, information that is `internally' meaningful, using a self-referentially
limited neural network model. Such a system has no {\it a priori} objects or laws, and is
evolved using a bootstrap system, so that it is the system itself that `internally' creates
patterns of relationships and their dominant modes of behaviour, and all (sub)systems are
fractal in character, that is, relationships within relationships, and so on {\it ad
infinitum}. In this way all emergent phenomena are unified, and it is this key feature that
has resulted in an understanding and linking, for the first time, of various phenomena. A key
feature of this process-physics is that this fracticality is associated with self-organising
criticality.

Previous progress in process physics
 included the demonstration that space and quantum physics are emergent  and
unified, with time  a distinct non-geometric process, that   quantum phenomena are caused by
fractal topological   defects  embedded in and forming a growing three-dimensional fractal
process-space, which is essentially a quantum foam.   Other   features of the emergent
physics were    quantum field theory with emergent flavour and confined colour, limited
causality and  the Born quantum measurement metarule,  inertia, time-dilation effects,
gravity and the equivalence principle, a growing universe with  a cosmological constant, 
black holes and event horizons, and the emergence of classicality.  

 As noted in     \cite{RC01} process physics results in
the emergence of a quantum-foam explanation for space, and for which quantum `matter'
effectively acts as a sink. This provides an explanation for the logical necessity of the
phenomenon of gravity.   Here, as well as updating previous results,  the
major new development is the derivation of General Relativity, along  with its  technical language
of  General Covariance.  General Relativity, and its key concept of {\it spacetime}, is
seen not to be fundamental but rather a  subtle phenomenological construct, arising as an
amalgam of two distinct phenomena: the `gravitational' characteristics of the emergent
quantum foam, and the classical `spacetime' measurement protocol, but with the later
violated by quantum measurements. As we shall see the `gravitational' effects of the quantum
foam causes  observers, using their measurement protocol,  to create a pseudo-Riemannian
manifold history book (the spacetime construct) of classical events,  but it is entirely
erroneous to think that the curvature of this construct is the {\it cause} of gravity, or in any way determines
how matter moves.     

A key long-standing goal of physics has been the unification of gravity and the quantum
theory. This unification could not be achieved within the old {\it non-process physics}, but 
has now been achieved with the new {\it process physics}, but as expected only at the expense
of  atleast one of the combatant theories, and it is general relativity that must concede
defeat. However the quantum theory has a shallow victory as it in turn is an emergent
phenomenon, arising from the non-geometric and non-quantum self-referentially-limited neural
network model which implements the semantic information approach to comprehending reality. 
Nevertheless it must be emphasised  that the conventional non-process physics models will
continue to be useful techniques in analysing problems, and in particular this applies to 
general relativity and the standard quantum theory.  What has been achieved in process physics
is the explanation of why reality must be so, and why the modes of behaviour are encodeable in
the syntax of the non-process physics. This older mode of dealing with reality will continue
to be used because for many problems it is eminently practical. It will require the
continued use of various metarules to overcome its limitations, but we now have an explanation
for them as well.

Process Physics shows that there  is a Quantum Gravity, but it is unrelated to General
Relativity and General Covariance, and essentially describes the emergent quantum phenomena
of the process-space or quantum foam, and its response to quantum `matter'. This Quantum
Gravity is manifested in the emergent Quantum Homotopic Field Theory for the process-space or
quantum foam, and  is logically prior to the emergence  of the general relativity
phenomenology, and cannot be derived from it.  It should be noted that as the general relativity 
syntax is, in part, the result of  the `classical' spacetime measurement protocol used by
observers, if one were considering the `quantising' of the classical phenomena of general
relativity, and  `quantisation' is always a  dubious enterprise at best, then one would sensibly
remove the observer protocol effects before undertaking such a task.  Unfortunately the
conventional orthodoxy has resulted in the total obscurification of  this insight.

The ongoing failure of physics to fully match all the aspects
of the phenomena of time, apart from that of order,  arises because physics has always
used non-process models, as is the nature of formal or syntactical systems. Such  systems
do not require any notion of process - they are entirely structural and static. The new
process physics overcomes these deficiencies by using a non-geometric
process model for time, but process physics  also  argues
for the importance of  relational or semantic information in modelling reality. Semantic information
refers to the notion that reality is a purely informational system where the information is
internally meaningful.    Hence the
information is `content addressable', rather than is the case in the usual syntactical
information modelling where the information is represented by symbols.  This symbolic or
syntactical mode is only applicable to higher level phenomenological descriptions, and 
for that reason was discovered first.

 A pure semantic information system must be  formed by
a subtle bootstrap process. The mathematical model for this has the form of a stochastic
neural network (SNN)  for the simple reason that neural networks are well known for their
pattern or non-symbolic information  processing abilities   \cite{neural}.  The stochastic
behaviour is related to the limitations of syntactical systems discovered by
G\"{o}del   \cite{G} and more recently extended  by
Chaitin   \cite{Chaitin90,Chaitin99,Chaitin01}, but also results in the neural network being
innovative in that it creates its own patterns.  The neural network  is self-referential,
and   the stochastic input,  known as self-referential noise, acts both to limit the depth
of the self-referencing   and also to generate potential order. 

This work reports on  the ongoing development of process physics beginning with a
 discussion of the comparison of the syntactical and the new semantic information system
and their connections with G\"{o}del's incompleteness theorem. Later sections describe
the emergent unification of gravitational and quantum phenomena, amounting to a quantum
theory of gravity. In particular the derivation of general relativity and its 
downgrading to a phenomenological tool is presented.

\section{Syntactical and Semantic Information Systems}

In modelling reality with  formal or syntactical information systems  physicists assume that
the full behaviour of a physical system can be compressed into axioms and rules for the
manipulation of symbols. However G\"{o}del discovered that self-referential syntactical
systems (and these include basic mathematics) have fundamental
limitations which amount to the realisation that not all truths  can be compressed into an
axiomatic structure, that formal systems are much weaker than previously supposed.  In physics
such systems have always been used in conjunction with metarules and metaphysical assertions, all
being `outside' the formal system and designed to overcome the limitations of the syntax. Fig.1
depicts the current understanding of self-referential syntactical systems. Here the key feature is
the G\"{o}del boundary demarcating the provable from the unprovable truths of some system. 
Chaitin, using Algorithmic Information Theory, has demonstrated that in mathematics the unprovable truths are
essentially random in character.  This, however, is a structural randomness in the sense that the individual truths
do not have any structure to them which could be exploited to  condense them down to or be encoded in
axioms. This is unlike random physical events which occur in time.  Of course syntactical
systems are based on the syntax of symbols and this is essentially non-process or
non-timelike.

\vspace{-10mm}

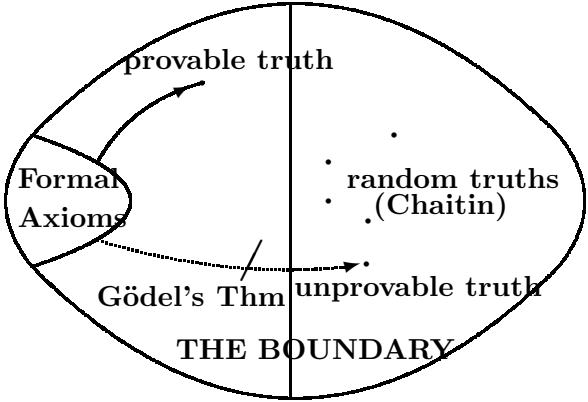
\begin{figure}[ht]
\hspace{17mm}
\setlength{\unitlength}{1.75mm}
\hspace{20mm}\begin{picture}(40,25)
\thicklines
\qbezier(0,10)(20,30)(40,10)
\qbezier(0,0)(20,-20)(40,0)
\qbezier(0,0)(-4,5)(0,10)
\qbezier(40,10)(44,5)(40,0)

\put(19.7,-10){\line(0,1){30}}

\qbezier(0,0)(15,5)(0,10)

\put(-1,6){{\bf Formal}}
\put(-1,3){{\bf  Axioms}}

\qbezier[60](5,2)(15,-1)(24,0)
\put(24,0){\vector(4,1){1}}
\put(16,-1){\line(1,2){1.5}}
\put(5.0,-3.0){\bf G\"{o}del's Thm}
\put(25.5,0.2){\circle*{0.5}}
\put(20,-2.2){\bf unprovable truth}

\put(13,14){\circle*{0.5}}

\put(7,15){\bf  provable truth}
\qbezier(5,8)(7.5,12.5)(13,14)
\put(12,13.71){\vector(2,1){0.1}}

\put(22.6,5){\circle*{0.5}}
\put(22.6,8){\circle*{0.5}}
\put(27.6,10){\circle*{0.5}}
\put(25.6,3.5){\circle*{0.5}}
\put(24,6){\bf random  truths}
\put(26,4.0){\bf (Chaitin)}

\put(11,-7){\bf THE BOUNDARY}

\end{picture}
\vspace{15mm}\caption{\small Graphical depiction of the `logic space' of
a self-referential syntactical information  system, showing the formal system
consisting of symbols and rules, and an example of one theorem
 (a provable truth).  Also shown  are unprovable truths which in
general are random (or unstructured) in character, following the work of Chaitin.
The G\"{o}delian boundary is the demarcation between provable
and unprovable truths. 
 \label{section:figure:Godel}}
\end{figure}

There is an analogy between the structure of self-referential syntactical
information systems and the present structure of quantum theory, as depicted in Fig.2.

\vspace{-10mm}

\begin{figure}[ht]
\hspace{17mm}
\setlength{\unitlength}{1.75mm}
\hspace{20mm}\begin{picture}(40,25)
\thicklines
\qbezier(0,10)(20,30)(40,10)
\qbezier(0,0)(20,-20)(40,0)
\qbezier(0,0)(-4,5)(0,10)
\qbezier(40,10)(44,5)(40,0)

\put(19.7,-10){\line(0,1){30}}

\qbezier(0,0)(15,5)(0,10)

\put(-1,6){{\bf Formal}}
\put(-1,3){{\bf  Axioms}}

\qbezier[60](5,2)(15,-1)(24,0)
\put(24,0){\vector(4,1){1}}
\put(8.0,0.0){\bf Born}
\put(7.0,-1.5){\bf Metarule}
\put(25.5,0.2){\circle*{0.5}}
\put(21,-2.2){\bf unprovable truth}

\put(13,14){\circle*{0.5}}

\put(7,15){\bf  provable truth}
\put(11,12){\bf (eg $E_n$)}
\qbezier(5,8)(7.5,12.5)(13,14)
\put(12,13.71){\vector(2,1){0.1}}

\put(22.6,5){\circle*{0.5}}
\put(22.6,8){\circle*{0.5}}
\put(27.6,10){\circle*{0.5}}
\put(25.6,3.5){\circle*{0.5}}
\put(24,6){\bf random  truths}

\put(13,-6){\bf A G\"{O}DEL }\put(12,-8){\bf BOUNDARY? }

\end{picture}
\vspace{15mm}\caption{\small Graphical depiction of the syntactical form of
conventional quantum theory.  The Born measurement metarule appears to bridge a
G\"{o}del-like boundary.
\label{section:figure:Quantum}}

\end{figure}
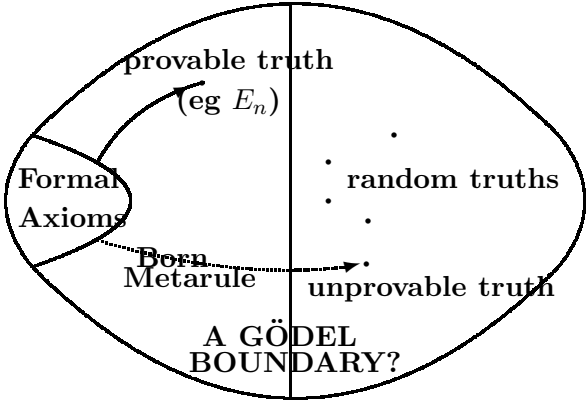

There the formal and hence non-process mathematical structure is capable of producing
many provable truths, such as the energy levels of the hydrogen atom, and these are also true in
the sense that they agree with reality.  But from the beginning of quantum theory the Born
measurement metarule was introduced to relate this non-process modelling to the actual randomness
of quantum measurement events.  The individuality of such random events is not a part of the
formal structure of quantum theory. Of course it is well known that  the non-process or 
structural aspects of the probability metarule  are consistent with the mathematical formalism, in
the form of the usual `conservation of probability' equation and the like.  Further, the  quantum
theory has always  been subject to various metaphysical interpretations, although these
have never played a key role for practitioners of the theory.  This all suggests that
perhaps the Born metarule is bridging a G\"{o}del-type boundary, that there is a bigger
system required to fully model quantum aspects of reality, and that the boundary is evidence of
self-referencing in that system.

Together the successes and failures of 
physics suggest that  a generalisation of  the traditional use of  syntactical 
information theory is required to model reality, and that this has now been identified as a
semantic information system which has the form of a stochastic neural network.

\begin{figure}[ht]
\vspace{-10mm}
\setlength{\unitlength}{1.75mm}

\hspace{40mm}\begin{picture}(40,25)
\thicklines
\qbezier(0,10)(20,30)(40,10)
\qbezier(0,0)(20,-20)(40,0)
\qbezier(0,0)(-4,5)(0,10)
\qbezier(40,10)(44,5)(40,0)


\put(20.1,-10){\line(0,1){14}}
\put(20.1,4.45){\line(0,1){5.6}}
\put(20.1,10.4){\line(0,1){3.6}}
\put(20.1,14.4){\line(0,1){5.6}}

\qbezier[70](0,0)(15,5)(0,10)
\qbezier(40,0)(25,5)(40,10)
\qbezier(37,-3)(18,5)(37,13)

\put(-1.5,6.0){{\bf Induced}}
\put(-2.0,4.35){{\bf  Formal}}
\put(-1.5,2.8){{\bf Axioms}}

\qbezier[60](5,2)(15,-1)(24,0)
\put(24,0){\vector(4,1){1}}
\put(16,-1.2){\line(1,2){1.2}}
\put(5.5,-3.0){\bf G\"{o}del's Thm}
\put(25,0.2){\circle*{0.5}}

\qbezier[20](7.4,4.3)(9,3)(14,4)
\put(13.28,3.8){\vector(4,1){1}}
\put(9,2.5){\line(1,2){1.2}}

\put(14,14.25){\oval(2,2)[t,l]}
\put(14,14.25){\oval(2,2)[b,l]}
\put(14,14.5){\oval(2,1.5)[t,r]}
\put(14,14.0){\oval(2,1.5)[b,r]}
\put(13.6,14.0){{\bf .}}\put(13.1,13.9){{\bf .}}
\put(13.7,14.6){{\bf .}}\put(13.4,13.5){{\bf .}}
\put(14.9,14.4){\line(1,0){5.25}}
\put(15,14){\line(1,0){5.15}}

\put(4,15.5){\bf ensemble truth}
\qbezier[70](5,8)(7.5,12.5)(12.8,14)
\put(12,13.71){\vector(2,1){0.1}}

\put(34,7.5){\vector(-3,1){20}}

\put(34,6.0){{\bf Stochastic}}
\put(34,4.25){{\bf Process}}
\put(34.5,2.5){{\bf System}}
\put(27.0,5.6){{\bf SOC}}
\put(26.7,4.02){{\bf syntax}}
\put(28,2.2){{\bf filter}}
\put(34,2.5){\vector(-1,-1){5}}\put(28.8,-2.7){\circle*{0.5}}
\put(21,-4.3){\bf random (contingent) }
\put(25,-5.9){\bf  truth}

\put(13,-7){\bf INDUCED}
\put(13,-9){\bf BOUNDARY}


\put(11,10.25){\oval(2,2)[t,l]}
\put(11,10.25){\oval(2,2)[b,l]}
\put(11,10.5){\oval(2,1.5)[t,r]}
\put(11,10.0){\oval(2,1.5)[b,r]}
\put(11.92,10.4){\line(1,0){8.25}}
\put(11.92,10){\line(1,0){8.25}}
\put(10.9,9.9){{\bf .}}\put(10.2,9.9){{\bf .}}
\put(10.5,10.5){{\bf .}}\put(10.5,9.5){{\bf .}}

\put(14,4.25){\oval(2,2)[t,l]}
\put(14,4.25){\oval(2,2)[b,l]}
\put(14,4.5){\oval(2,1.5)[t,r]}
\put(14,4.0){\oval(2,1.5)[b,r]}
\put(14.1,3.5){{\bf .}}\put(13.1,4.49){{\bf .}}
\put(13.6,4.5){{\bf .}}\put(13.85,3.9){{\bf .}}
\put(14.9,4.5){\line(1,0){5.2}}
\put(14.9,4){\line(1,0){5.25}}

\put(25,5){\circle*{0.5}}
\put(23,10){\circle*{0.5}}
\put(27,12){\circle*{0.5}}
\put(23,15){\circle*{0.5}}
\put(29,16){\circle*{0.5}}
\end{picture}
\vspace{15mm}
\caption{\small Graphical depiction of the bootstrapping of and the emergent structure of a
self-organising pure semantic information system.  As a high level effect  we see the
emergence of an induced formal system, corresponding to the current standard syntactical
modelling of reality. There is an emergent  G\"{o}del-type boundary which represents the 
inaccessibility of the random or contingent truths from the induced formal or syntactical 
system.  A process of self-organised criticality (SOC)  filters out the seeding or bootstrap
  syntax.    
\label{section:figure:Process}}
\end{figure}
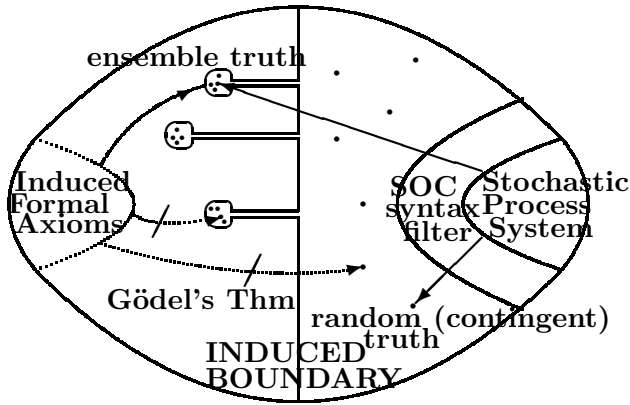

Fig.3 shows a graphical depiction of the bootstrapping of a pure 
semantic information  system, showing the stochastic    neural network-like process
system from which the semantic system is seeded or bootstrapped.  Via a
Self-Organised Criticality Filter (SOCF) this seeding  system is removed or hidden.
  From the process system,  driven by
Self-Referential Noise (SRN), there are  emergent
truths, some of which are generically true (ensemble truths)  while others are 
purely contingent.  The
ensemble truths are also reachable from the
Induced Formal System as theorems, but from which, because of the non-process nature of the
induced formal system,  the contingent truths cannot be reached. In this manner there arises  a
G\"{o}del-type boundary.  The existence of the latter leads to    induced metarules 
that enhance  the
induced formal system, if that is to be used solely  in higher order phenomenology.

Western science and philosophy has always been dominated by non-process thought.  This `historical record'
or  {\it being} model of reality has been with us since
Parmenides, and his student Zeno, of Elea, and is known as the Eleatic model (c500 BCE). However, nevertheless, and for the
dubious reason of generating support for his supervisors {\it being} model of reality,  Zeno gave us the
first insights into the inherent problems of comprehending motion, a problem long forgotten by conventional
non-process physics, but finally explained by process physics. The {\it becoming} or {\it processing} model of 
reality dates back  to Heraclitus of Ephesus (540-480 BCE) who argued that  common sense is mistaken in
thinking that the world consists of stable things; rather the world is in a state of flux. The appearances of
`things' depend upon this flux for their continuity and identity.  What needs to be explained, Heraclitus argued,
is not change, but the appearance of stability. With process physics western science and philosophy is now able to
move beyond the moribund non-process mindset.  While it was the
work of G\"{o}del who demonstrated beyond any doubt that the non-process system of thought had fundamental
limitations; implicit in his work is that the whole reductionist mindset that goes back to Thales of Miletus could
not offer, in the end, an effective account of reality. However the notion that there were limits to syntactical or
symbolic encoding is actually very old. Priest    \cite{Priest} has given an account of that history.  However in the
East the Buddhists in particular were amazingly advanced in their analysis and comprehension of reality.  
Stcherbatsky
   \cite{BLogic}, writing about the extraordinary achievements of Buddhist logic in the C6 and C7$^{th}$ CE, noted
that; 
\begin{quote} {\it Reality according to Buddhists is
kinetic, not static, but logic, on the
other hand, imagines a reality
stabilized in concepts and names. The
ultimate aim of Buddhist logic is to
explain the relation between a moving
reality and the static constructions of
logic.} 
\end{quote}

In the West the process system approach to reality  was developed, much later, by such {\it process
philosophers} as Peirce, James, Bergson and Whitehead to name a few, although their achievements were very limited
and substantially flawed, limited as they were by the physical phenomena known to them.   A collection
of their writings is available in    \cite{ProcessPhilosophy}. Perhaps a quote from Charles Peirce  
\cite{ProcessPhilosophy}, writing in 1891, gives the sense of their thinking;

\begin{quote}

{\it The one intelligible theory of the universe is that
of objective idealism, that matter is effete mind, inveterate
habits becoming physical laws. But before this can be
accepted it must show itself capable of explaining the
tridimensionalty of space, the laws of motion, and the
general characteristics of the universe, with mathematical
clearness and precision; for no less should be demanded of
every philosophy.}
\end{quote}

With process physics we have almost achieved this end, and 
Wheeler has already expressed this notion of {\it inveterate habits} as ``law without law"
   \cite{Wheeler}. As the astute reader will note the self-referentially limited neural network model, that underpins
process physics, is remarkably akin to Peirce's {\it effete mind}. But it is the limitations of syntax, and the
need for intrinsic or semantic information `within' reality and at all levels, that reality is not imposed,   that
drives us to this approach.   Einstein, the modern day eleatic thinker, realised all too well the limitations of
non-process thinking, but was unable to move out of the non-process realm that the West had created for itself, 
for according to Carnap   \cite{Carnap};

 \begin{quote}
{\it Once Einstein said that the problem of the Now worried him seriously. He
explained that the
experience of the Now means something special for man, something
essentially different from the
past and the future, but that this important difference does not and cannot
occur within physics.
That this experience cannot be grasped by science seems to him a matter of
painful but inevitable resignation. I remarked that all that occurs
objectively can be described in science: on the one hand the temporal
sequence of events is described in physics; and, on the other hand, the
peculiarities of man's experiences with respect to time, including his
different attitude toward past, present and  future, can be described and
(in principle) explained in psychology. But Einstein thought that
scientific descriptions cannot possibly satisfy our human needs; that there
is something essential about the Now which is just outside of the realm of
science.}
\end{quote}

\section{Self-Referentially Limited  Neural Networks}
\label{sect:NN}

Here we briefly describe a model for a  self-referentially limited neural network and in the
following section we describe how such a network results in emergent quantum behaviour, and which,
increasingly, appears to be a unification of space and quantum phenomena. Process physics is  a semantic
information system and is devoid of {\it a priori} objects and their laws  and so it requires a subtle
bootstrap mechanism to set it up. We use a stochastic neural network, Fig.4a, having the structure of 
  real-number valued connections or relational information strengths $B_{ij}$ (considered as forming a
square matrix) between  pairs of nodes or pseudo-objects
$i$ and $j$. In standard 
neural networks   \cite{neural} the network  information resides in both link and node
variables,  with the semantic information residing in attractors of the iterative network.
Such systems are also not pure in that there is an assumed underlying and manifest {\it a
priori} structure.
  
 The nodes and their link variables  will be revealed  to be themselves sub-networks of informational
relations. To avoid explicit self-connections
$B_{ii}\neq 0$, which are a part of the sub-network content of
$i$, we use antisymmetry $B_{ij}=-B_{ji}$ to conveniently ensure that 
$B_{ii}=0$, see Fig.4b.

At this stage we are using a syntactical system with symbols $B_{ij}$ and, later, rules for the
changes in the values of these variables. This system is the syntactical seed for the pure
semantic system.   Then to ensure that the nodes and links are not remnant {\it a priori}
objects the system must generate strongly linked  nodes (in the sense that the $B_{ij}$ for
these nodes are much larger than the $B_{ij}$ values for non- or weakly-linked nodes) forming
a fractal network; then self-consistently the start-up nodes and links may themselves be
considered   as mere names for sub-networks of relations.  For a successful suppression  the scheme must
display self-organised criticality (SOC)   \cite{SOC} which acts as a filter for the start-up syntax. The
designation `pure' 
 refers to the notion that all seeding syntax has been removed. SOC is the process
where the emergent behaviour  displays universal criticality in that the  behaviour is
independent of the individual start-up syntax; such a start-up syntax  then  has no 
ontological significance.  

To generate a fractal structure we must  use a non-linear iterative system for the
$B_{ij}$ values.  These iterations amount to the  necessity to introduce a time-like
 process.   Any system possessing {\it a priori}  `objects' can never
be fundamental as the explanation of such objects must be outside  the system.  Hence in
process physics the absence of intrinsic undefined objects is linked with the phenomena of
time, involving as it does an ordering of `states', the present moment effect, and the
distinction between past and present. Conversely in non-process physics the presence of {\it a
priori } objects is related to the use of the non-process geometrical model of time, with this modelling
and its geometrical-time metarule being an approximate emergent description from process-time.  In this
way process physics arrives at a new modelling of time, {\it process time}, which is much more complex
than that introduced by Galileo, developed by Newton, and reaching its high point with Einstein's
spacetime geometrical model.  Unlike these geometrical models process-time does model  the {\it Now} effect. 
Process physics also shows that time cannot be modelled by any other structure, other than a time-like process, here an
iterative scheme. There is nothing {\it like} time available for its modelling.  The near obsession of theoretical
physicists with the geometrical modelling of time, and its accompanying notion of {\it determinism},  has done much to
retard the development of physics.  The success of process physics implies that time along with self-referencing is in some
sense prior to the other phenomena, and certainly prior to space, as will be seen in sect.7 within the discussion of a
multi-component universe.  
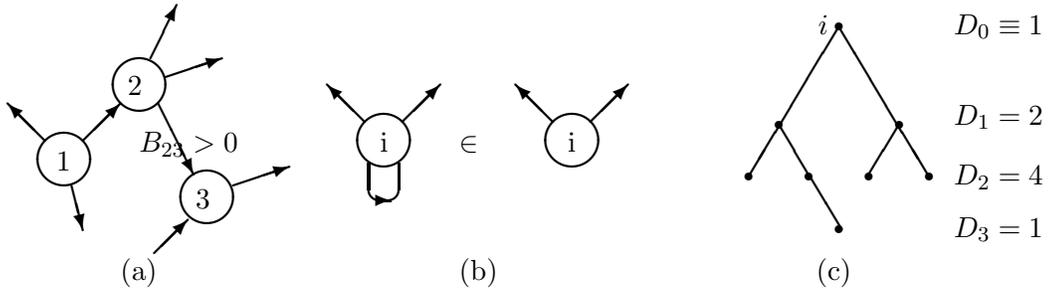
\begin{figure}
\setlength{\unitlength}{0.5mm}
\hspace{20mm}
\begin{picture}(150,60)
\thicklines
\put(10,10){\circle{15}}\put(8,7){1}
\put(30,30){\circle{15}}\put(27,27){2}
\put(48,0){\circle{15}}\put(45,-3){3}
\put(15,15){\vector(1,1){10}}\put(30,12){$B_{23}>0$}
\put(5,15){\vector(-1,1){10}}
\put(12,3){\vector(1,-4){3}}
\put(35,25){\vector(1,-2){9.5}}
\put(33,37){\vector(1,2){7}}
\put(37,32){\vector(3,1){15}}
\put(55,3){\vector(3,1){15}}
\put(34,-15){\vector(1,1){9}}
\put(95,15){\circle{15}}\put(94,12){i}
\put(100,20){\vector(1,1){10}}
\put(90,20){\vector(-1,1){10}}
\put(25,-22){(a)}   \put(115,-22){(b)}    \put(210,-22){(c)}

\put(145,15){\circle{15}}\put(144,12){i}
\put(150,20){\vector(1,1){10}}
\put(140,20){\vector(-1,1){10}}
\put(95,9){\oval(8,20)[b]}
\put(96,-1){\vector(1,0){1.0}} 
\put(115,12){$\in$}
\end{picture}
\hspace{40mm}
\setlength{\unitlength}{0.20mm}

\hspace{105mm}
\begin{picture}(0,50)(40,0)  
\thicklines
\put(155,165){\line(3,-5){60}}
\put(155,165){\line(-3,-5){60}}
\put(115,100){\line(3,-5){42}}
\put(195,100){\line(-3,-5){21}}
\put(135,160){ \bf $i$}
\put(225,160){ \bf $D_0\equiv 1$}
\put(225,100){ \bf $D_1=2$}
\put(225,60){ \bf $D_2=4$}
\put(225,25){ \bf $D_3=1$}
\put(155,165){\circle*{5}}
\put(115,100){\circle*{5}}
\put(195,100){\circle*{5}}
\put(95,65){\circle*{5}}
\put(135,65){\circle*{5}}
\put(175,65){\circle*{5}}
\put(215,65){\circle*{5}}
\put(155,30){\circle*{5}}
\end{picture}

\caption{\small (a) Graphical depiction of the neural network with links
$B_{ij}\in {\cal R}$ between nodes or pseudo-objects. Arrows indicate sign of
$B_{ij}$. (b) Self-links are internal to a node, so $B_{ii}=0$. (c) An $N=8$ spanning
tree for a random graph (not shown) with $L=3$.  The  distance distribution $D_k$ is indicated for
node {\it i}.
\label{section:figure:neural}}
\end{figure}

The  stochastic neural network   so far has been realised with one
particular  scheme involving a stochastic non-linear matrix iteration, see (\ref{eq:1}). 
The matrix inversion $B^{-1}$ then models self-referencing in that it requires  all
elements of $B$ to compute any one element of $B^{-1}$. As well there is the 
additive SRN  
$w_{ij}$ which limits the self-referential information  but, significantly, also acts in such
a way that the network is innovative in the sense of generating semantic information, that is
information which is internally meaningful.  The emergent behaviour is believed to be
completely generic in that it is not suggested that reality is a computation, rather it
appears that reality has the form of an self-referential order-disorder
information system.  It is important to note that process physics is a non-reductionist
modelling of reality; the basic iterator (\ref{eq:1}) is premised on the general assumption  that
reality is sufficiently complex that self-referencing occurs, and that this has limitations.
Eqn.(\ref{eq:1}) is then a minimal bootstrapping implementation of these notions.  At higher emergent levels
this self-referencing manifests itself  as {\it interactions} between emergent patterns, but other novel effects may
also arise.    

To be a successful contender for the Theory of Everything (TOE) process
physics must ultimately prove the uniqueness conjecture:  that the  characteristics (but not
the contingent details) of the  pure  semantic information system are unique.  This would
involve demonstrating both the effectiveness of the SOC filter and the robustness of the
emergent phenomenology, and the complete agreement of the later with observation.    

The stochastic neural network is  modelled by the iterative process\footnote{While not directly relevant it may be
helpful to the reader to outline the line of thought that led to (1), arising as it did from the quantum field
theory frontier of quark physics.  A highly effective approximation to QCD was developed that made extensive use of bilocal
fields and the functional integral calculus (FIC), see \cite{GCM} for reviews of this Global Colour Model (GCM).  The core
effect   in the GCM, which uses the Euclidean metric, but with analytic continuation back to the Minkowski metric, is the
constituent quark effect, which is a non-linear equation for those non-zero bilocal fields about which the induced
effective action for hadronic fields is to be expanded (the Dyson-Schwinger equation for the bilocal fields effective
action).  If we strip away the spacetime and quantum number indices from that equation, we arrive at (1), but without the
SRN term, as done in R.T. Cahill and C.M. Klinger, {\it     Pregeometric Model of the Spacetime Phenomenology}, Phys. Lett. 
A{\bf 223}(1996)313. But as is apparent there it was impossible to recover the quantum phenomena.  However if in the GCM the
constituent quark equations are extended by incorporating a bilocal stochastic term and so making the equation an
iterative one, then the iterations of this generalised equation generate not only the constituent quark effect but all the
emergent hadronic field phenomenology.  This is because of the mysterious stochastic quantisation procedure discovered by
G. Parisi and Y. Wu, {\it Scientia Sinica} {\bf 24}, 483(1981). This stochastic noise was interpreted as the new intrinsic
Self-Referential Noise in \cite{CK97}, when the connection with the work of G\"{o}del and Chaitin became apparent, and we
finally arrive at (1). Hence beneath quantum field theory there is evidence of a self-referential stochastic neural network,
and its interpretation as a semantic information system. Only by discarding the spacetime background of QFT do we discover
the necessity for time, space, and the quantum.}
\begin{equation}
B_{ij} \rightarrow B_{ij} -\alpha (B + B^{-1})_{ij} + w_{ij},  \mbox{\ \ } i,j=1,2,...,2M;
M
\rightarrow
\infty,
\label{eq:1}\end{equation}
 where 
 $w_{ij}=-w_{ji}$ are
independent random variables for each $ij$ pair and for each iteration and chosen from some probability
distribution. Here $\alpha$ is a parameter the precise value of which should not be critical but which
influences the self-organisational process. 
We start the iterator at 
$B\approx 0$, representing the absence of information.  
  With the noise absent the iterator 
behaves in a deterministic and reversible manner, giving a condensate-like system, given by
the matrix
\begin{equation}
B =  MDM^{-1}; \mbox{\ \ \ \  }D=\left(\begin{array}{rrrrrr}
0 & +b_1 & 0 & 0\\
-b_1 & 0 & 0 & 0 \\
0 & 0 & 0 & +b_2 \\
0 & 0 & -b_2 & 0 & \\
 &&&&. \\ &&&&&. \\          
\end{array}\right),  \mbox{\ \ \ \ \  }  b_1,b_2,... \geq 0,
\label{eq:2}\end{equation}
where $M$ is  a real orthogonal matrix determined uniquely by the start-up $B$, and each $b_i$
evolves according to the iterator
$
b_i \rightarrow b_i-\alpha(b_i-b_i^{-1})
$, but it does serve as the non-metriciseable background for the muti-world universe discussed later. In the presence
of the noise the iterator process  is non-reversible and non-deterministic.  It   is also manifestly
non-geometric and non-quantum, and so does not assume any of the standard features of syntax based
physics models.   The dominant  mode is  the  formation of an apparently randomised 
background (in
$B_{ij}$) but, however, it  also manifests  a   self-organising process which results
in a growing  three-dimensional  fractal process-space that competes with this random background -
the formation of a `bootstrapped universe'.  Here we report on the current status of ongoing
work to extract the nature of this `universe'. 

The emergence of order in this system might
appear to violate expectations regarding the 2nd Law of Thermodynamics; however because of the SRN
the system behaves as an open system and the  growth of order arises from the
self-referencing term, $B^{-1}$ in (\ref{eq:1}), selecting certain implicit order in the SRN.  Hence the SRN
acts as a source of negentropy\footnote{The term {\it negentropy} was introduced by E.
Schr\"{o}dinger   \cite{negentropy} in 1945, and since then there has been ongoing discussion of its 
meaning. In process physics it manifests as the SRN.}

 This  growing  three-dimensional  fractal process-space is an example of a Prigogine far-from-
equilibrium dissipative structure   \cite{Prigogine} driven by the SRN. 
 From each iteration the noise term
will additively introduce rare large value
$w_{ij}$.  These  $w_{ij}$, which define sets of strongly linked nodes, will persist   through more
iterations than smaller valued $w_{ij}$ and, as well,  they become further linked  by the iterator
to form a three-dimensional process-space with embedded topological defects. In this way the
stochastic neural-network creates stable strange attractors and as well determines their interaction
properties. This information is all internal to the system; it is the semantic information within
the network.
 
 To see the nature of this internally generated information  consider a node $i$ involved in one
such large
$w_{ij}$;   
 it will be   connected via  other large $w_{ik}$ to a
number of other nodes and so on, and this whole set of connected nodes forms a connected random graph unit
which we call a gebit as it acts as a small piece or bit of geometry formed from random information links and 
from which the process-space is self-assembled. The gebits compete for new links  and undergo
mutations. Indeed, as will become clear, process physics is remarkably analogous in its operation to
biological systems. The reason for this is becoming clear: both reality and subsystems of reality must use
semantic information processing to maintain existence, and symbol manipulating systems are totally unsuited
to this need, and in fact totally contrived.

\newpage

.

\vspace{-40mm}

\hspace{-20mm}\includegraphics{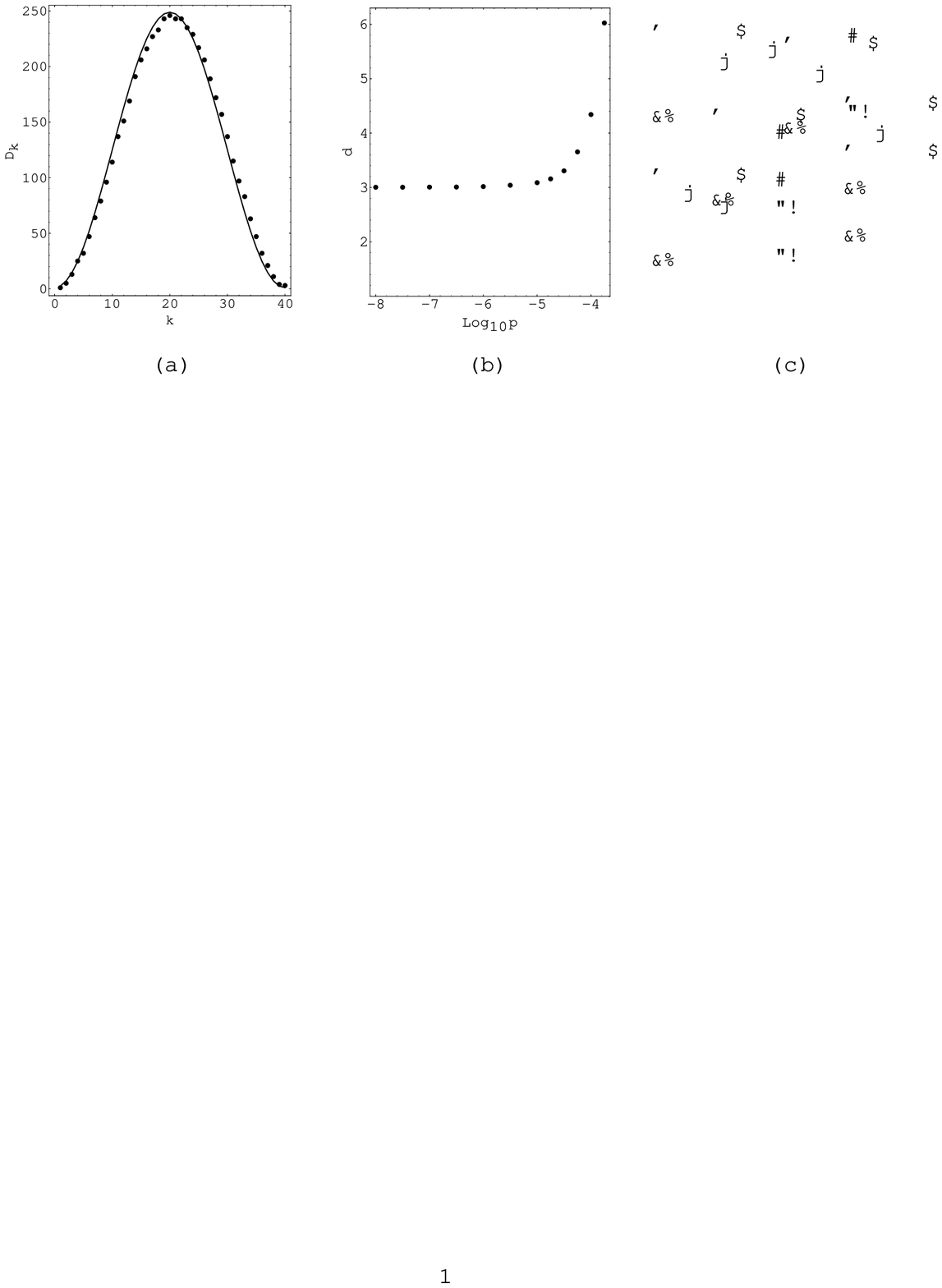}

\vspace{70mm}
\begin{figure}[ht]
\caption{\small 
(a) Points show the $D_k$ set and $L=40$ value found by numerically  maximising ${\cal P}[D,L,N]$
for $\mbox{Log}_{10}p=-6$ for fixed  $N=5000$. Curve shows
$D_k\propto \sin^{d-1}(\frac{\pi k}{L})$ with best fit $d=3.16$ and $L=40$, showing  excellent
agreement, and indicating  embeddability in an $S^3$ with some topological defects. (b) Dimensionality $d$ of
the gebits as a function of  the probability $p$. (c) Graphical depiction of the `process space' at one stage of the iterative
process-time showing a quantum-foam structure formed from  embeddings and linkings of gebits.  The
linkage connections have the distribution of a 3D space, but the individual gebit components are
closed compact spaces and cannot be embedded in a 3D background space.  So the drawing is only
suggestive. Nevertheless this figure indicates that process physics generates a cellular
information system, where the behaviour is determined at all levels by internal information. } 
\end{figure}

To analyse the connectivity of such 
gebits assume for simplicity that the large $w_{ij}$  arise with fixed but very small probability $p$,
then the geometry of  the gebits is revealed by studying the probability distribution for  the structure
of the random graph units or gebits minimal spanning trees with $D_k$ nodes  at $k$ links from node $i$
($D_0
\equiv 1$), see Fig.4c, this is given by   \cite{Nagels}

\begin{equation}{\cal P}[D,L,N] \propto \frac{p^{D_1}}{D_1!D_2!....D_L!}\prod_{i=1}^{L-1}
(q^{\sum_{j=0}^{i-1}{D_j}})^{D_{i+1}}(1-q^{D_i})^{D_{i+1}},\label{eq:3}\end{equation}
where $q=1-p$, $N$ is the total number of nodes in the gebit and $L$ is the maximum depth from node $i$. 
To find the most likely connection pattern we numerically maximise ${\cal P}[D,L,N]$ for fixed $N$ with respect
to
$L$ and the $D_k$. The resulting $L$ and $\{D_1,D_2,...,D_L\}$ fit very closely to the form $D_k\propto
\sin^{d-1}(\pi k/L)$;  see Fig.5a  for $N=5000$ and $\mbox{Log}_{10}p=-6$.  The resultant  $d$
values for a range of $\mbox{Log}_{10}p$ and $N=5000$ are shown in Fig.5b.

This shows, for $p$ below a critical value, that
$d=3$, indicating that the connected nodes have a natural embedding in a 3D hypersphere $S^3$;
call this a base gebit. Above that value of $p$,   the increasing value of $d$
indicates the presence of extra links that, while some conform with the embeddability,  are in the main defects
with respect to the geometry of the $S^3$.  These extra links act as topological defects.  By themselves these
extra links will have the   connectivity and embedding geometry of numbers of gebits, but these gebits have a
`fuzzy' embedding in the base gebit. This is an indication of  fuzzy  homotopies  (a homotopy is,
put simply, an embedding of one space into another).   Here we see the emergence of geometry, not only of space but the
internal symmetry spaces of quantum fields.

The base gebits $g_1, g_2, ...$ arising from the SRN together with their embedded topological defects have
another remarkable property:  they are `sticky' with respect to the iterator.  Consider the   larger valued 
$B_{ij}$ within a given gebit  $g$, they form  tree graphs and most tree-graph adjacency matrices are singular
 (det($g_{tree})=0$).  However  the presence of other smaller valued $B_{ij}$ and the
general
 background noise  ensures that det$(g)$ is small  but not exactly zero. 
Then the 
$B$ matrix has an inverse with large components that act to cross-link  the new and
existing gebits. This cross-linking is itself random, due to the presence of background noise, and the above
analysis  may again be used and we would conclude that  the base gebits themselves are formed into a 3D hypersphere
with embedded topological defects.  The nature of the resulting 3D process-space is suggestively indicated in Fig.5c, and
behaves essentially as a quantum foam   \cite{foam1}.

Over ongoing
iterations the existing gebits become cross-linked and eventually lose their ability to undergo further
linking; they lose their `stickiness'  and decay. The value of the parameter $\alpha$ in (\ref{eq:1}) must
be small enough that the `stickiness' persists over many iterations, that is, it is not quenched
too quickly, otherwise the emergent network  will not grow.  Hence the emergent
space is 3D but is continually undergoing  replacement of its component gebits;  it is an
informational process-space, in sharp distinction to the non-process continuum geometrical spaces
that have played a dominant role in modelling physical space.  If the noise is `turned off' then
this emergent dissipative space will decay and cease to exist.  We thus see that the nature of
space is deeply related to the logic of the limitations of logic, as implemented here as a
self-referentially limited neural network.

\section{Modelling Gebits and their Topological Defects}

We need to extract convenient but approximate syntactical descriptions  of the semantic information in the
network, and these will have the form of a sequence of  mathematical constructions, the first being the Quantum
Homotopic Field Theory. Importantly they must all retain explicit manifestations of the SRN. To this  end first
consider the special case of the iterator when the SRN is frozen at a particular
$w$, that is we consider iterations with  an artificially  fixed SRN term. Then the iterator is equivalent to the
minimisation of an `energy' expression (remember that
$B$ and $w$ are antisymmetric)
\begin{equation}
E[B;w]= -\frac{\alpha}{2}\mbox{Tr}[B^2]-\alpha \mbox{TrLn}[B]+\mbox{Tr}[wB].
\label{eq:4}\end{equation}
Note that for disconnected gebits $g_1$ and $g_2$ this energy is additive, $E[g_1\oplus g_2]=E[g_1]+E[g_2]$.
Now suppose the fixed $w$ has the form of a  gebit approximating an $S^3$ network with one embedded topological
defect which is itself an $S^3$ network, for simplicity.  So we are dissecting the gebit into base gebit, defect gebit
and linkings or embeddings between the two. We also ignore the rest of the network, which is permissible if our gebit
is disconnected from it.  Now  if det$(w)$ is not small, then this gebit is non-sticky, and for small $\alpha$, the
iterator converges to $B\approx\frac{1}{\alpha}w$, namely an enhancement only of the gebit.  However because the
gebits are rare constructs they tend to be composed of larger $w_{ij}$ forming tree structures, linked by smaller
valued
$w_{ij}$.  The tree components make det$(w)$ smaller, and then the inverse $B^{-1}$ is activated and generates new
links.  Hence, in particular, the topological defect relaxes, according to
the `energy' measure, with respect to the base gebit.  This relaxation  is an example of a `non-linear elastic'
process   \cite{Ogden}.  The above gebit has the form of  a mapping  $\pi: S \rightarrow \Sigma$ from a base space to a
target space.  Manton   \cite{Manton1, Manton2, GP} has constructed the continuum form for the `elastic energy' of such
an embedding and for $\pi: S^3 \rightarrow S^3$  it is the Skyrme energy
\begin{equation}
E[U]=\int \left[ -\frac{1}{2}\mbox{Tr}(\partial_i UU^{-1}\partial_i UU^{-1}) -\frac{1}{16} \mbox{Tr}[\partial_i
UU^{-1},\partial_i UU^{-1}]^2\right],
\label{eq:5}\end{equation}
where $U(x)$ is an element of $SU(2)$. Via the parametrisation $U(x)=\sigma(x)
+i\vec{\pi}(x).\vec{\tau}$, where the
$\tau_i$ are Pauli matrices,  we have $\sigma(x)^2+\vec{\pi}(x)^2$=1, which parametrises an $S^3$ as
a unit hypersphere  embedded in $E^4$ (which has no ontological significance, of course). Non-trivial minima of
$E[U]$ are known as Skyrmions (a form of topological soliton), and have
$Z=\pm1,\pm2,...$, where $Z$ is the winding number of the map, 
\begin{equation}
Z=\frac{1}{24\pi^2}\int\sum\epsilon_{ijk}\mbox{Tr}(\partial_i UU^{-1}\partial_j UU^{-1}\partial_k
UU^{-1}).\label{eq:6}
\end{equation}

The first key to extracting  emergent phenomena from the stochastic neural network is the validity of this continuum
analogue, namely that $E[B;w]$ and $E[U]$ are describing essentially the same `energy'
reduction process.  This should be amenable to detailed  analysis.  

This `frozen' SRN analysis of course does not match the time-evolution of the
full iterator (\ref{eq:1}), for  this displays a much richer collection of processes.  With ongoing new noise
in each iteration and the saturation of the linkage possibilities of the gebits emerging from this noise, there
arises a process of ongoing birth, linkaging and then decay of most patterns.  The task is then to
identify those particular patterns that survive this flux, even though all components of
these patterns eventually disappear, and to attempt a description of their modes of
behaviour.  This brings out the very biological nature of the information processing in
the SNN, and which appears to be characteristic of a `pure' semantic information system.  Hence the emergent `laws of
physics' are the habitual habits of this system, and it appears that they maybe identified.  However there is no encoding
mechanism for these `laws', they are continually manifested; there is no {\it cosmic code}.  In contrast living or
biological systems could be defined as those emergent patterns which discovered how to encode their `laws', in a
syntactical {\it genetic code}. Nevertheless such biological systems make extensive use of semantic information at all
levels, as their genetic code is expressed in the phenotype.

In general each gebit, as it emerges from the SRN, has active nodes and embedded topological defects, again with
active nodes.  Further there will be defects embedded in the defects and so on,  and so gebits begin to have the
appearance of a fractal defect structure, and all the defects having various classifications and associated  winding
numbers.  The energy analogy above suggests that  defects with opposite  winding numbers at the same fractal depth 
may annihilate by drifting together and merging. Furthermore the embedding of the defects is unlikely to be
`classical', in the sense of being described by a mapping
$\pi(x)$, but rather would be fuzzy, i.e described by some functional, $F[\pi]$, which would correspond to a
classical embedding only if $F$ has a very sharp supremum at one particular $\pi=\pi_{cl}$. As well these gebits are
undergoing linking because their active nodes (see    \cite{CK98} for more discussion) activate the $B^{-1}$ new-links
process between them, and so by analogy the gebits themselves form larger structures with embedded fuzzy
topological defects. This emergent behaviour is suggestive of a quantum  space foam, but one containing
topological defects which will be preserved by the system, unless annihilation events occur.  If these
topological defects are sufficiently rich in fractal structure as to be preserved, then their initial
formation would have occurred as the process-space relaxed out of its initial, essentially random form.
This phase would correspond to the early stages of the Big-Bang. Once the topological defects are trapped
in the process-space  they are doomed to meander through that space by essentially self-replicating, i.e.
continually having their components die away and be replaced by similar components.  These residual
topological defects are what we call matter. The behaviour of both the process-space and its defects is
clearly determined by the same network processes; we have an essential unification of space and matter
phenomena.  This emergent quantum foam-like behaviour suggests that the full generic  description of the
network behaviour  is via the Quantum Homotopic Field Theory (QHFT) of the next section.  We also see that cellular
structures are a general feature of semantic information systems, with the information necessarily distributed.

\section{Modelling the Emergent Quantum Foam}

To construct this QHFT we introduce an appropriate configuration space, namely  all the possible homotopic 
mappings $\pi_{\alpha\beta}: S_\beta \rightarrow S_\alpha$, where the 
$S_1,S_2,..$, describing `clean' or topological-defect free gebits, are compact spaces of various types. Then  QHFT
has the form of an iterative functional Schr\"{o}dinger equation for the discrete time-evolution of a wave-functional 
$\Psi[....,\pi_{\alpha\beta},....;t]$
\begin{equation}\Psi[....,\pi_{\alpha\beta},....;t+\Delta
t]= \Psi[....,\pi_{\alpha\beta},....;t]
-iH\Psi[....,\pi_{\alpha\beta},....;t]\Delta t +  \mbox{QSD terms}.
\label{eq:7}\end{equation}  
This  form arises as it is models the preservation of semantic information, by means of a unitary time
evolution, even in the presence of the noise in the QSD terms.  We thus see the origin of the Hilbert space
structure of quantum phenomena.  Because of the QSD noise  (\ref{eq:7}) is an irreversible quantum system.   
     The time step $\Delta t$ in (\ref{eq:7})  is relative to the scale of the
fractal processes being explicitly described, as we are using a configuration space of mappings between
prescribed gebits.  At smaller scales we would need a smaller value of   $\Delta t$.  Clearly this invokes a
(finite) renormalisation scheme. We now discuss the form of the hamiltonian and the Quantum State Diffusion
(QSD) terms.

First (\ref{eq:7}), without the QSD term,  has a form analogous to  a `third quantised' system, in
conventional terminology   \cite{Baby}. These 
systems were considered   as   perhaps capable of generating  a quantum theory of
gravity. The argument here  is that this is the emergent behaviour of the SNN, and it does indeed lead to quantum
gravity, but with quantum matter as well.  More importantly we understand the origin of (\ref{eq:7}), and it
will lead to quantum and then classical gravity, rather than arise from classical gravity via some ad hoc or
heuristic quantisation procedure.

Depending on the `peaks' of 
$\Psi$ and the connectivity of the resultant dominant mappings such mappings are to be interpreted as
either embeddings  or links; Fig.5c then suggests the dominant process-space form within
$\Psi$ showing both links and embeddings. The emergent process-space then has the characteristics of a  
quantum foam. Note that, as indicated in Fig.5c, the original start-up links and nodes are now absent.
Contrary to the suggestion in Fig.5c, this process space cannot be embedded  in a {\it finite}
dimensional geometric space with the emergent metric preserved, as it is composed of  nested
or fractal finite-dimensional closed spaces.  

We now consider the form of the hamiltonian H.  The previous section suggested that Manton's non-linear
elasticity interpretation of the Skyrme energy is appropriate to the SNN.  This then suggests that H is
the functional operator
\begin{equation}
H=\sum_{\alpha\neq\beta}h[\frac{\delta}{\delta
\pi_{\alpha\beta}},\pi_{\alpha\beta}],
\label{eq:8}\end{equation}
where $h[\frac{\delta}{\delta \pi},\pi]$ is the (quantum) Skyrme Hamiltonian functional operator for the system based
on  making fuzzy the  mappings
$\pi: S \rightarrow \Sigma$, by having $h$ act on wave-functionals of the form $\Psi[\pi(x);t]$. Then $H$
is the sum of pairwise  embedding or homotopy hamiltonians. The corresponding functional Schr\"{o}dinger
equation would simply  describe the time evolution of quantised Skyrmions with the base space fixed, and
$\Sigma \in SU(2)$. There have been very few analyses of even this class of problem, and then the base space
is usually taken to be $E^3$.  We shall not give the explicit form of $h$ as it is complicated, but wait to present
the associated action. 

In the absence of the QSD terms the time evolution in (\ref{eq:7}) can be formally written as a functional
integral
\begin{equation}
\Psi[\{\pi_{\alpha\beta}\};t']=\int\prod_{\alpha\neq\beta}{\cal
D}\tilde{\pi}_{\alpha\beta}e^{iS[\{\tilde{\pi}_{\alpha\beta}\}]}
\Psi[\{\pi_{\alpha\beta}\};t],
\label{eq:9}\end{equation}
where, using the continuum $t$ limit notation, the action is a sum of pairwise actions,
\begin{equation}
S[\{\tilde{\pi}_{\alpha\beta}\}]=\sum_{\alpha\neq\beta}S_{\alpha\beta}[\tilde{\pi}_{\alpha\beta}],
\label{eq:10}\end{equation}
\begin{equation}
S_{\alpha\beta}[\tilde{\pi}]=\int_t^{t'}dt''\int d^nx\sqrt{ -g} \left[ \frac{1}{2}\mbox{Tr}(\partial_\mu
\tilde{U}\tilde{U}^{-1}\partial^\mu
\tilde{U}\tilde{U}^{-1}) +\frac{1}{16} \mbox{Tr}[\partial_\mu \tilde{U}\tilde{U}^{-1},\partial^\nu
\tilde{U}\tilde{U}^{-1}]^2\right],
\label{eq:11}\end{equation}
and the now time-dependent (indicated by the tilde symbol) mappings $\tilde{\pi}$ are parametrised by
$\tilde{U}(x,t)$, $\tilde{U}\in S_\alpha$. The metric $g_{\mu\nu}$ is that of the $n$-dimensional base space,
$S_\beta$, in
$\pi_{\alpha,\beta}: S_\beta
\rightarrow  S_\alpha$. As usual in the functional integral formalism the functional derivatives in the quantum
hamiltonian, in (\ref{eq:8}), now manifest as the time components $\partial_0$ in (\ref{eq:11}),  so now
(\ref{eq:11}) has the form of a `classical' action, and we see the emergence of `classical' fields, though the
emergence of `classical' behaviour is a more complex  process.  Eqns.(\ref{eq:7}) or (\ref{eq:9}) describe an
infinite set of quantum skyrme systems, coupled in a  pairwise manner.  Note that each homotopic mapping
appears in both orders; namely
$\pi_{\alpha\beta}$ and 
$\pi_{\beta\alpha}$.

The   Quantum State Diffusion (QSD)   \cite{QSD} terms  are non-linear and
stochastic, 
\begin{equation}
\mbox{QSD terms} =\sum_\gamma\left(
<\!\!L^\dagger_\gamma\!\!>L_\gamma-\frac{1}{2}L^\dagger_\gamma L_\gamma-
<\!\!L^\dagger_\gamma\!\!><\!\!L_\gamma\!\!>\right)
\Psi\Delta t+\sum_\gamma\left(L_\gamma-<\!\!L_\gamma\!\!> \right)\Psi\Delta \xi_\gamma,
\label{eq:12}\end{equation}
which involves summation over the class of Linblad functional operators $L_\gamma$.
The QSD terms  are  up to 5th order in $\Psi$, as in general,
\begin{equation}
<\!\! A \!\!>_t=\int \prod_{\alpha\neq\beta}{\cal D}\pi_{\alpha\beta}\Psi[\{
\pi_{\alpha\beta} \};t]^* A 
\Psi[\{ \pi_{\alpha\beta}\};t],
\label{eq:13}\end{equation}
and where $\Delta \xi_\gamma$
are complex statistical variables with means  $M (\Delta \xi_\gamma) = 0$, $M( \Delta
\xi_\gamma
\Delta\xi_{\gamma'})= 0$  and $M(\Delta \xi^*_\gamma\Delta\xi_{\gamma'}) =
\delta(\gamma-\gamma')\Delta t$

These QSD terms are ultimately responsible for the emergence of
classicality via an  objectification process, but in particular  they produce
wave-function(al) collapses during quantum measurements, as the QSD terms tend to `sharpen' the fuzzy homotopies
towards  classical or sharp  homotopies (the forms of the Linblads will be discussed in detail elsewhere).   So the
QSD terms, as residual SRN effects, lead to 
 the Born quantum measurement  random behaviour, but here arising from the process physics, and not being
invoked as a metarule.
 Keeping the QSD terms leads to a functional integral representation for a
density matrix formalism in place of (\ref{eq:9}), and this amounts to a derivation of  the decoherence
formalism which is usually arrived at by invoking the Born measurement metarule. Here we see that
`decoherence'  arises from the limitations on self-referencing.

In the above we have  a deterministic and unitary
evolution, tracking and preserving topologically encoded information, together with the stochastic
QSD terms, whose form protects that information during localisation events, and which
also  ensures the full matching in QHFT of process-time to real time: an ordering of
events, an intrinsic direction or `arrow' of time and a modelling of the contingent
present moment effect.   So we see that process physics generates a complete theory of quantum
measurements involving the  non-local, non-linear and  stochastic QSD terms.  It does this because
it generates both the `objectification' process associated with the classical apparatus and the actual
process of (partial)  wavefunctional collapse as the quantum modes interact with the measuring
apparatus.  Indeed many of the mysteries of quantum measurement are resolved when it is realised that
it is the measuring apparatus itself that actively provokes the collapse, and it does so because the
QSD process is most active when the system deviates strongly from its dominant mode, namely the
ongoing relaxation of the system to a 3D process-space, and matter survives only because of its topological form.  This is
essentially the process that Penrose   \cite{Penrose} suggested, namely that the quantum measurement process is essentially
a manifestation of quantum gravity. The demonstration of the validity of the Penrose argument of course
could only come about when  quantum gravity was {\it derived} from deeper considerations, and not by
some {\it ad hoc} argument such as the {\it quantisation} of Einstein's classical spacetime model.

The mappings  $\pi_{\alpha\beta}$ are related to group manifold parameter spaces with the group determined by
the dynamical stability of the mappings. This symmetry leads to  the flavour symmetry of the
standard model. Quantum homotopic mappings or skyrmions  behave as fermionic\footnote{However in order to establish fermionic behaviour a Wess-Zumino process
must be extracted from the iterator behaviour or the QHFT. Such a WZ process is  time-dependent, and so cannot arise
from the frozen SRN argument in Sect.4.} or bosonic modes for 
appropriate winding numbers; so process physics predicts both fermionic and bosonic quantum modes, but
with these  associated with topologically encoded information and not with  objects or `particles'. 

\section{Quantum Field Theory}
The QHFT is a very complex `book-keeping' system for the emergent properties of the neural network,
and we now sketch how we may extract a more familiar quantum field theory (QFT) that relates to the
standard model of `particle' physics. An effective QHFT should reproduce the emergence
of the process-space part of the quantum foam, particularly its 3D aspects. The QSD processes play
a key role in this as they tend to enhance classicality. Hence at an appropriate scale QHFT
should approximate to a more conventional QFT, namely the emergence of a wave-functional system
$\Psi[U(x);t]$ where the  configuration space is that of homotopies from a 3-space to $U(x) \in
G$, where $G$ is some group manifold space. This $G$
describes `flavour' degrees of freedom.  So we are coarse-graining out the 
gebit structure of the quantum-foam. 
 Hence the Schr\"{o}dinger
wavefunctional equation for this QFT  will have the form
\begin{equation}\Psi[U;t+\Delta
t]= \Psi[U;t]
-iH\Psi[U;t]\Delta t +  \mbox{QSD terms},
\label{eq:14}\end{equation} 
where the general form of $H$ is known, and where a new residual manifestation of the SRN appears
as the new QSD terms.  This system describes skyrmions embedded in a continuum space. It is
significant that such Skyrmions are only stable, at least in flat space and for static skyrmions, if
 that space is 3D.  This tends to confirm the observation that 3D space is special for the neural
network process system.
Again, in the absence of the QSD terms, we may express (\ref{eq:15}) in terms of the functional integral
\begin{equation}
\Psi[U;t']=\int{\cal
 D}\tilde{U}e^{iS[\tilde{U}]}
\Psi[U;t].
\label{eq:15}\end{equation}
To gain some insight into the phenomena present in (\ref{eq:14}) or (\ref{eq:15}), it is convenient to use the
fact that functional integrals of this Skyrmionic form my be written in terms of  Grassmann-variable
functional integrals, but only by introducing a fictitious `metacolour' degree of freedom and 
associated coloured fictitious  vector bosons. This is essentially the reverse of the Functional Integral Calculus
(FIC) hadronisation technique in  the Global Colour Model (GCM) of QCD   \cite{GCM}.  The action for the Grassmann and
vector boson  part of the system is of the form (written for flat space)
\begin{equation}
S[\overline{p},p,A^a_\mu]=\int d^4x\left( \overline{p}\gamma^\mu(i\partial_\mu
+g\frac{\lambda^a}{2}A^a_\mu)p-\frac{1}{4}F^a_{\mu\nu}(A)F^{a\mu\nu}(A)
\right),
\label{eq:16}\end{equation}
where the  Grassmann variables $p_{f c}(x)$ and $\overline{p}_{f c}(x)$ have
flavour and metacolour labels. The Skyrmions are then re-constructed, in this system, as  topological solitons formed from
the low energy  Nambu-Goldstone modes; other emergent modes are of higher energy and can be
ignored.  These coloured and flavoured but fictitious fermionic fields
$\overline{p}$ and $p$  correspond to the proposed preon system   \cite{Preons1,Preons2}. As
they are purely fictitious, in the sense that there  are no excitations in the system corresponding
to them, the metacolour degree of freedom must be hidden or confined.  We thus arrive at the general feature of the 
standard model of particles with flavour anf confined colour degrees of freedom. Then while the QHFT and the QFT represent
an induced  syntax for the semantic information, the preons may be considered as an induced `alphabet' for that syntax. The
advantage of introducing this preon alphabet is that we can more easily determine the states of the system by using  the
more familiar  language of fermions and bosons, rather than working with the skyrmionic system, so long as only
colour singlet states are finally permitted.  However it is important to note that (\ref{eq:16}) and the action in
(\ref{eq:15}) are certainly not the final forms.  Further analysis will be required to fully extract the induced
actions for the emergent QFT.  

\section{Inertia and Gravity}
Process physics predicts that the neural network behaviour will be characterised by a growing
3-dimensional process-space having, at a large scale, the form of a $S^3$ hypersphere, which is
one of the forms allowed by Einstein's syntactical modelling. It is possible to give the dominant
rate of growth of this hypersphere. However first, from  random graph theory   \cite{randomg}, we
expect more than one such spatial system, with 
each having the character of a growing hypersphere, and all embedded in the random background
discussed previously. This background has no metric structure, and so these various hyperspheres
have no distance measure over them. We have then a multi-world universe (our `universe' being
merely one of these `worlds'). Being process spaces they compete for new gebits, and so long as we
avoid a saturation case, each will grow according to
\begin{equation}
\frac{dN_i}{dt}=aN_i-bN_i  \mbox{\ \ \ \ \ } a>0, b>0,
\label{eq:17}\end{equation}
where the 
last term describes the decay of gebits at a rate $b$, while the first describes growth of the
$i^{\mbox{th}}$ `world', this  being proportional to the size  (as measured by its gebit content  number)
$N_i(t)$ as success in randomly attaching new gebits is proportional to the number of gebits 
present (the `stickiness' effect), so long as we ignore the topological defects (quantum `matter')
as these have a different stickiness, and also affect the decay rate, and so slow down the
expansion. Thus
$N_i(t)$ will show exponential growth, as appears to be the case as indicated by recent observations
of very distant supernovae counts   \cite{cosmo}. Hence process physics predicts a 
positive  cosmological constant, and that this is unrelated to the phenomenon of gravity.  Indeed this
multi-world model is incompatible with general relativity, as it is not capable of even describing the
non-geometric background or embedding system. In this enlarged cosmology each world would have its own
independent big-bang beginning, but then it is no longer necessary for this on-going ensemble of worlds to
have had a beginning itself, as we may presumably take the system start-up time to $-\infty$.   Hence the observation of the
cosmological constant effect is here to be interpreted as arising from the existence of these other worlds.   

One striking outcome of process physics is an explanation for the phenomenon of gravity. 
First note that matter is merely topological defects embedded in the process space, and we expect
such defects to have a larger than usual gebit connectivity; indeed matter is a violation of the
3-D connectivity of this space, and it is for this reason we needed to introduce fields to emulate
this extra non-spatial connectivity.  One consequence of this is that in the region of these
matter fields the gebits decay faster, they are less sticky because of the extra connectivity. 
Hence in this region, compared to other nearby matter-free regions the gebits are being `turned
over' more frequently  but at the same time are less effective in attracting new gebits. Overall
this suggests that  matter-occupying-regions act as net sinks for gebits, and there will be a
trend for the neighbouring quantum foam to undergo a diffusion/relaxation process in which
the foam effectively moves towards the matter: matter acts as a sink for space, but never as a
source. Such a process would clearly correspond to gravity. As  the effect is described by
a net diffusion/relaxation of space which  acts equally on all surrounding matter, the
in-fall mechanism is independent of the nature of the surrounding matter. This is nothing
more than Einstein's Equivalence Principle.   As well if the in-fall rate exceeds the rate at which
`motion' through the process-space is possible then an event horizon appears, and this  is clearly  the
black hole scenario.  Presumably at the core of a black-hole is a tangle of topological defects, so that the
effective dimensionality is very high, and which maintains the in-flow of quantum foam. Then here the General
Relativity formalism fails.

Finally we mention one long standing unsolved problem in physics, namely an understanding of
inertia. This is the effect where objects continue in uniform motion  unless acted upon by `forces',
and was first analysed by Galileo in the modern era, but of course Zeno made an early attempt. However there has never been
an explanation for this effect; in Newtonian physics it was built into the syntactical description rather than being a
prediction of that modelling.  Of course current physics is essentially a static modelling of reality, with
motion indirectly accessed via the geometrical-time metarule, and so the failure to explain motion
is not unexpected. However process physics offers a simple explanation.

The argument for inertia follows from a simple self-consistency argument. Suppose a 
topological defect, or indeed a vast bound collection of such defects, is indeed `in motion'. This
implies that the gebits are being preferentially replaced  in the direction of that `motion', for
motion as such is a self-replication process; there is no mechanism in process physics for a fixed
pattern to say `slide' {\it through} the quantum-foam.  Rather motion is self-replication of the
gebit  connectivity patterns in a set direction.  Since the newest gebits, and hence the stickiest
gebits, in each topological defect,  are on the side corresponding to  the direction of motion, the
 gebits on that side are preferentially replaced.  Hence the motion is self-consistent and
self-perpetuating.

An additional effect expected in process physics is that such motion results in a time dilation and length
contraction effects;  the self-replication effect  is to be considered as partly  the self-replication 
associated with any internal oscillations and partly self-replication associated with `motion'. This
`competition for resources' results in the slowing down of internal oscillations, an idea discussed by 
Toffoli   \cite{Toffoli}.   Such effects have been seen in a variety of `non-relativistic'  continuum
systems \cite{AU}, and indeed they have a long history. In particular emergent Lorentz symmetry has been analysed in the
modelling of dislocations in crystals   \cite{HG},  where `breather modes' as solitons  arise.  Hence the lesson
is that emergent Lorentz symmetry is generic whenever there is a unification of the substratum system and embedded
excitations, here    the soliton as a dynamical  emergent feature within some dynamical background, rather than being
merely `pasted' onto some {\it a priori} geometrised and so structureless background.   Bell   \cite{Bell} has
argued for this dynamical interpretation of Lorentz symmetry, as indeed did H.A. Lorentz himself, until this view was
overwhelmed by the Einstein interpretation of this symmetry. This is discussed in more detail in the next section. 
  More recently similar ideas have
emerged  \cite{BU,MV} in the analysis of the sound waves in classical fluids.   Here we assume that the QHFT will also
display these generic Lorentzian dynamical effects, namely the time-dilation/length-contraction effects.

 Geometry is clearly an emergent but approximate phenomenological language; it is certainly not fundamental, and the
implicit assumption in physics that it is fundamental has caused many problems and confusions.

There is one further novel effect that is of some significance. The  quantum-foam appears to
represent a special frame of reference, but one that is well hidden by the time-dilation and length
contraction effects. Hence we would not expect a Michelson-Morley type experiment  to reveal this frame. 
However using the arguments of  Hardy   \cite{Hardy} and Percival   \cite{IP}  we  expect that
the action of the QSD wavefunctional `collapse' processes would reveal the actual frame 
through a multi-component   EPR experiment, as the non-local QSD collapse occurs in a truly
simultaneous manner.   By not using the non-local quantum collapse process we are essentially deceived  by the subtle
effects of the quantum foam dynamics, as we shall now see. 
 
\section{General Relativity}

It is essentially straightforward to derive the formalism of general relativity from the above considerations
of the behaviour of the quantum foam, with the key steps following the work of  Kirkwood in
1953    \cite{RK1, RK2}; see also the recent work by Martin    \cite{TM}. Assuming that the diffusion/relaxation of
the quantum foam in the presence of matter may be described, at the classical level, by a flow field with velocity
${\bf v}({\bf r},t)$, and that some classical test object has velocity ${\bf v}_0(t)=d{\bf r}(t)/dt$, where ${\bf
r}(t)$ is the position of the object, hence some cosmological coordinate system is invoked\footnote{Some discussion of this is
in \cite{RK1, TM}, though  it requires much more analysis.  Possibly the non-local connections in the
QHFT  provide some preferred frame, coinciding wth the CBR isotropic frame. The time
$t$ in (\ref{eq:18}),  (\ref{eq:19}), etc is presumably the  iterator time, since this is also  the time
that indicates non-local simultaneous wavefunctional collapses.  So it appears that the process
$t-r$ frame may be  defined by quantum non-locality effects.}.  Then the key measure is the  velocity of
the object relative to the foam,
\begin{equation}
{\bf v}_R(t) ={\bf v}_0(t) - {\bf v}({\bf r}(t),t),
\label{eq:18}\end{equation} 
where the foam flow may be explicitly time-dependent. 
The key assertion is then  that the change in this relative  velocity, in the presence  of a
non-gravitational force $F$, is given by \cite{RK1},
\begin{equation}
\frac{d}{dt}\left(m({\bf v}_R){\bf v}_R\right)=F-m({\bf v}_R)({\bf v}_R.{\bf  \nabla}).{\bf v}({\bf r},t),
\label{eq:19}\end{equation}
 which is non-linear as  
\begin{equation}
m({\bf v}_R)=\frac{m_0}{\surd(1-\frac{{\displaystyle \bf v}_R^2}{\displaystyle c^2})},
\label{eq:20}\end{equation}
is the usual speed-dependent effective mass from the Lorentzian dynamics for an object with rest mass $m_0$,  which
goes along with time dilations and length contractions caused by effective `motion' through the quantum foam.
In keeping with the equivalence principle $m_0$ cancels from the equation, when $F=0$, and we consider only this case in the
following. Here
$c$ is the maximum speed of any disturbance through the quantum foam, and is clearly relative to the foam. The effective
force on the RHS of (\ref{eq:19}) follows from   the usual fluid mechanics expression for the acceleration of the fluid.
 Note that the denominator in (\ref{eq:20}) immediately leads to event horizon
effects wherever $|{\bf v}|=c$, where the region where $|{\bf v}|< c$ is inaccessible from the region where $|{\bf v}| > c$. 

As Kirkwood showed the  general relativity formalism of  test object trajectories essentially follows from
(\ref{eq:19}) even though it has no notion of curvature or even of general covariance. The only real curvature  from
process-physics is the cosmological curvature, and this is not related to gravity.  As we shall see the curvature of
general relativity actually arises from introducing the classical `spacetime' measurement protocol, and so is potentially
misleading though not without practical use.  This protocol also results in the formalism of general covariance.   The key
dynamics is not that of the fictitious spacetime metric but rather the non-linear dynamics of the quantum flow, and in this
regard the successes of general relativity should be regarded as providing valuable clues regarding the foam.  We shall now
show from (\ref{eq:19}) and the Lorentz effects how the language of general relativity (GR) arises. 
Eqn.(\ref{eq:19}) may be written as

\begin{equation}
\frac{d{\bf v}_R(t)}{dt}=-({\bf v}_R({\bf r},t).{\bf  \nabla }).{\bf v}({\bf r},t)-\frac{1}{m}\frac{dm}{dt}{\bf
v}_R(t).
\label{eq:21}\end{equation}
Evaluating the $dm/dt$ term and solving for $d{\bf v}_R/dt$,  we obtain
\begin{equation}
\frac{d{\bf v}_R(t)}{dt}=-({\bf v}_R(t).{\bf  \nabla }).{\bf v}({\bf r},t)+\frac{1}{c^2}
({\bf v}_R(t).({\bf v}_R({\bf r},t).{\bf  \nabla }){\bf v}({\bf r},t)){\bf v}_R(t),
\label{eq:22}\end{equation} which is cubic in ${\bf v}_R$.  The event horizon effect is now more subtle.
 Now (\ref{eq:21}) or (\ref{eq:22}) is nothing more than the Euler-Lagrange equation from minimising the  functional
\begin{equation}
\tau[{\bf v}_0,{\bf v}]=\int dt\surd(1-\frac{{\bf v}_R^2}{c^2})
\label{eq:23}\end{equation}
with respect to ${\bf r}(t)$, for  the case of an  irrotational flow  $\nabla \times {\bf v}={\bf 0}$.
Eqn.(\ref{eq:23}) looks very non-GR involving as it does a velocity relative to the flowing foam and the speed $c$
also relative to the foam.  However (\ref{eq:23}) gives
\begin{equation}
d\tau^2=dt^2-\frac{1}{c^2}(d{\bf r}(t)-{\bf v}({\bf r}(t),t)dt)^2,
\label{eq:24}\end{equation}
which is  the   Panlev\'{e}-Gullstrand form of the metric   \cite{PP, AG} for GR restricted to the object
trajectory, and is very suggestive.  However to fully recover the GR formalism we need to explicitly introduce
the spacetime measurement protocol, and the peculiar effects that it induces for the observers historical records.

The quantum foam, it is argued, induces actual dynamical time dilations and length contractions in agreement
with the Lorentz interpretation of special relativistic effects.  Then observers in  uniform motion
`through' the foam will, on measurement say of the speed of light, obtain always the numerical value  $c$.   To see this
explicitly consider how various observers $O, P, Q,..$, in different  relative motions through the foam, measure the speed
of light.  They  each acquire a standardised rod of length $l_0$, and then  measure the time $\Delta t_0$, using an
accompanying standardised clock, for a light pulse to travel to one end and be reflected back to the starting end.  The
observed speed of light is defined for any observer using
$c_P=$distance travelled/time duration. For an observer and rod at rest wrt the foam, we get $c_P=c$. However the
real Lorentzian dynamics mean that the length of the moving rod is reduced to $l=l_0\surd(1-v_R^2/c^2)$ and its clock
time is reduced to
 $\Delta t=\Delta t_0\surd(1-v_R^2/c^2)$, and then  $c_P=2l/\Delta t=2l_0/\Delta t_0=c$.  Hence the Lorentzian
dynamics results in all observers assigning the same speed $c$ to light; using classical protocols they are unaware of
the underlying Lorentzian dynamics at work.
The classical spacetime measurement protocol of the observers is based on this effect, namely that the observed speed of
light is always $c$, and that even  observers in relative motion  agree on this one value.

To be explicit the measurement protocol actually exploits this subtle effect by using the radar method for
assigning historical spacetime coordinates to an event: the observer records the time of emission and reception
of radar pulses ($t_r > t_e$), and then retrospectively assigns the time and distance of an event $B$ according
to ( ignoring directional information for simplicity) 
\begin{equation}T_B=\frac{1}{2}(t_r+t_e), \mbox{\ \ \ } D_B=\frac{c}{2}(t_r-t_e),\label{eq:25}\end{equation} 
where each observer is now using the same numerical value of $c$.
 The event $B$ is then plotted as a point in 
an individual  geometrical construct by each  observer,  known as a spacetime record, with coordinates $(D_B,T_B)$. This
is no different to a historian recording events according to  some agreed protocol.  Unlike historians, who don't
confuse history books with reality, physicists do so. We now show that, because of this protocol and the quantum foam
dynamical effects, observers will discover, on comparing their historical records of the same events, that the expression

\begin{equation}
 \tau_{AB}^2 =   T_{AB}^2- \frac{1}{c^2} D_{AB}^2
\label{eq:26}\end{equation}
is an invariant, where $T_{AB}=T_A-T_B$ and $D_{AB}$ are the difference in times and distances assigned to events $A$
and
$B$ using the measurement protocol (\ref{eq:25}), so long as both are sufficiently small compared with the scale
of inhomogeneities  in the velocity field.  We now demonstrate this for the situation illustrated in Fig.6. For observer
$O$:  
\begin{equation}\tau_{AB}^2=T_B^2=t_B^2(1-\frac{v_R^2}{c^2}),
\label{eq:27}\end{equation}

\begin{figure}[t]
\hspace{17mm}
\setlength{\unitlength}{1.5mm}
\hspace{30mm}\begin{picture}(40,45)
\thicklines
\put(-2,-2){{\bf $A$}}
\put(24,21){\bf $B$ $(T_B)$}
\put(25,-3){\bf $r$}
\put(20,10){\bf flow  }\put(26,10.5){\vector(1,0){10}}
\put(28,8.5){\bf $v>0$}
\put(-5,30){\bf $t$}
\put(34,31){ O($v_0$)}
\put(16,45){ O$^\prime(v_0^\prime$)}

\put(0,0){\line(1,0){45}}
\put(0,0){\line(0,1){50}}
\put(0,50){\line(1,0){45}}
\put(45,0){\line(0,1){50}}

\put(0,0){\line(1,1){35}}
\put(0,0){\line(1,3){16}}

\put(1.5,17.5){\bf $(T_e^\prime)$}
\put(4.0,30.5){\bf $(T_B^\prime)$}\put(9.5,30.5){\line(1,0){0.9}}
\put(6.7,16.5){\bf $e$}
\put(10,43){\bf $(T_r^\prime)$}
\put(16,43){\bf $r$}
\put(6,18){\vector(4,1){16}}
\put(22,22){\vector(-1,3){7.3}}
\put(20,33){\bf $c-v$}
\put(12,21){\bf $c+v$}

\put(-3,17.5){\bf $t_e$}\put(-0.5,18){\line(1,0){0.9}}
\put(-3,22){\bf $t_B$}\put(-0.5,22){\line(1,0){0.9}}
\put(-3,43){\bf $t_r$}\put(-0.5,43){\line(1,0){0.9}}

\put(13,18){$\gamma$}
\put(16,33){$\gamma$}

\end{picture}
\caption{\small  A process-time - process-space $t-r$ representation (showing only one spatial direction) 
of two observers $O$ and $O^\prime$ moving in the direction of the effective flow of the quantum foam,
with speeds $v_0$ and $v_0^\prime$ with respect to (wrt) the frame $r$, and so with speeds $v_R=v_O-v$ and
$v_R^\prime=v_0^\prime-v$, wrt to the foam. The diagram is drawn for the case $v>0$.  The flow is assumed to
be uniform over the scale of these events. Event
$A$ is when both observers pass, and is also used to define zero time and zero distance for all systems, for
convenience.  Observer
$O^\prime$ emits a pulse of light ($\gamma$) at event $e$, which propagates through the foam with speed $c$,
but with speed $c+v$ wrt to the
$t-r$ frame, and  which is reflected from observer $O$ at event $B$, and propagates back to $O^\prime$
against the flow with speed $c-v$, and is received by $O^\prime$ at event $r$.  Times in brackets
$(\mbox{\ })$ are the protocol times assigned to various events by the observers.  For example $T^\prime_B$ is the
time of event $B$ assigned by $O^\prime$ using the spacetime measurement protocol, Eqn.(\ref{eq:25}). As
discussed in the text, the measurement protocol results in such observers introducing a pseudo-Riemannian curved
spacetime construct to record and relate the history of such restricted `classical'  events.  In this way
the protocol induces the syntax of general covariance.}
\end{figure}

\noindent since $T_A=0$ and $D_{AB}=0$, while for observer $O^\prime$:
\begin{equation}
\tau^{\prime 2}_{AB}=T^{\prime 2}_B-\frac{1}{c^2}D^{\prime 2}_B
=\frac{1}{4}(T^\prime_r+T^\prime_e)^2-\frac{1}{4}(T^\prime_r-T^\prime_e)^2=T^\prime_rT^\prime_e=t_rt_e
(1-\frac{v^{\prime 2}_R}{c^2}).
\label{eq:28}\end{equation}
Now from Fig.6 we note the distance relationships
\begin{equation}
(c+v)(t_B-t_e)=v_ot_B-v_0^\prime t_e,  \mbox{\ \ \  } (c-v)(t_r-t_B)=v_ot_B-v_0^\prime t_r,
\label{eq:29}\end{equation}
from which we obtain
\begin{equation}
t_rt_e=\frac{1-
\frac{\displaystyle v_R^2}{\displaystyle c^2}}{1-\frac{\displaystyle v^{\prime 2}_R}{\displaystyle c^2}}t_B^2,
\label{eq:30}\end{equation}
and we see, from (\ref{eq:27}), (\ref{eq:28}) and (\ref{eq:30}),  that $ \tau_{AB}^2= \tau^{\prime 2}_{AB}$. 
Hence the protocol and Lorentzian effects result in the construction in (\ref{eq:26})  being indeed an
invariant.  This  is  a remarkable and subtle result.  For Einstein this invariance was a fundamental
assumption, but here it is a derived result. Explicitly indicating  small quantities  by $\Delta$ prefixes,
and on comparing records retrospectively, an ensemble of nearby observers  agree on the invariant
\begin{equation}
\Delta \tau^2=\Delta t^2-\frac{1}{c^2}\Delta d^2,
\label{eq:31}\end{equation} 
for any two nearby events.  This implies that their individual patches of spacetime records may be mapped one
into the other merely by a change of coordinates, and that collectively the spacetime patches  of all may
be represented by one pseudo-Riemannian manifold, where the choice of coordinates for this manifold is
arbitrary, and we finally arrive at the invariant 
\begin{equation}
\Delta\tau^2=g_{\mu\nu}(x)\Delta x^\mu \Delta x^\nu.
\label{eq:inv}\end{equation} Of course this spacetime construction loses the experiential time
effects, and is indeed a static non-processing historical record, and is nothing more than a very refined
history book, with the shape of the manifold encoded in a metric tensor $g_{\mu\nu}(x)$.  However this encodes
two quite distinct processes, the velocity flow field of the quantum foam and as well the measurement protocol.
Enormous confusion enters if we do not keep these two aspects in mind. Of course it is utterly nonsensical to
assert that this spacetime construct {\it is} reality.   In general (\ref{eq:inv}) may be reduced to the
Panlev\'{e}-Gullstrand form (\ref{eq:24}), but now not restricted to events on the trajectory of an  object, that is to
space-like events (this requires further constructive analysis similar to that in Fig.6). By in part requiring
agreement with Newtonian gravity  Hilbert and Einstein guessed the equation which specifies how matter, and energy-momentum
in general, determines $g_{\mu\nu}(x)$, 
\begin{equation}
R_{\mu\nu}-\frac{1}{2}g_{\mu\nu}R=-8\pi G T_{\mu\nu},
\label{eq:32}\end{equation}
where $R_{\mu\nu}$ and $R$ are contractions with $g_{\mu\nu}$ of the curvature tensor
$R_{\mu\nu\lambda\kappa}$, and 
$T_{\mu\nu}$ is the  energy-momentum tensor, in the usual manner.  This has been tested in a number of
situations including, in particular, the indirect evidence of wave phenomena  from the studies of binary
pulsars.  Such gravitational waves are merely the quantum-foam flow pattern adjusting to changes in the
 sink locations and velocities. However if we remove the measurement protocol effects, and the
Panlev\'{e}-Gullstrand form of the metric does this,  then the Einstein equation (\ref{eq:32}) should really be
regarded as indicating the quantum foam flow dynamics, and how the presence of  energy-momentum, that is,
topological defects, acting as sinks for the foam,  influences that flow. A task of process-physics, using the
QHFT, is to derive that dynamics, but the GR equation (\ref{eq:32}) clearly can provide valuable clues.  See
\cite{RK2} for early speculation.  The cosmological constant, which arises from a non-gravitational process
(Sect.7.) presumably can be included in (\ref{eq:32}), in a phenomenological manner.

This differential geometry system can be used to determine the trajectories of test objects using the familiar
form
\begin{equation}
\Gamma^\lambda_{\mu\nu}\frac{dx^\mu}{d\tau}\frac{dx^\nu}{d\tau}+\frac{d^2x^\lambda}{d\tau^2}=0
\label{eq:33}\end{equation}
which involves the usual affine connection of the spacetime construct, and which arises from minimising the
functional
\begin{equation}
\tau[x]=\int dt\sqrt{g^{\mu\nu}\frac{dx^{\mu}}{dt}\frac{dx^{\nu}}{dt}},
\label{eq:path}\end{equation}
wrt to the path $x[t]$.  Now (\ref{eq:33}) must predict the same trajectory as the quantum-foam form
(\ref{eq:19}).   It is instructive to consider the special
case of
 a spherically symmetric mass $M$ as the sink for the quantum foam.  In the absence of a derivation of the
quantum-foam dynamics from the QHFT, like Einstein, we shall assume agreement with Newtonian gravity in the
limit of  object speeds low relative to $c$. Then we can neglect the last term in (\ref{eq:21}), and
(\ref{eq:21}) leads to Newton's equation of motion 
\begin{equation}
\frac{d^2 {\bf r}(t)}{dt^2}=-GM\frac{\hat{\bf r}}{r^2},
\label{eq:34}\end{equation} 
only if the velocity field is the static in-ward flow
\begin{equation}
{\bf v}({\bf r})=-\sqrt{\frac{2GM}{r}}\hat{\bf r}.
\label{eq:35}\end{equation}
Then with this velocity field the Panlev\'{e}-Gullstrand metric becomes, and now being now valid for events not
on the trajectory of the object, and with spherical symmetry,
\begin{equation}
d\tau^2=dt^2-\frac{1}{c^2}(dr+\sqrt{\frac{2GM}{r}}dt)^2-\frac{1}{c^2}r^2d\theta^2,
\label{eq:36}\end{equation}
which, under  the change of coordinate,
\begin{equation}
t^\prime=t-
\frac{2}{c}\sqrt{\frac{2GMr}{c^2}}+\frac{4GM}{c^2}\mbox{tanh}^{-1}\sqrt{\frac{2GM}{c^2r}},
\label{eq:37}\end{equation}
we obtain the Schwarzschild form of the metric
\begin{equation}
d\tau^2=(1-\frac{2GM}{c^2r})dt^{\prime 2}-
\frac{r^2d\theta^2}{c^2}-\frac{dr^2}{c^2(1-\frac{\displaystyle 2GM}{\displaystyle c^2r})},
\label{eq:38}\end{equation}
which is a particular solution of (\ref{eq:32}).
Hence using this form of the metric and (\ref{eq:33}) we obtain the same trajectory predictions as from 
(\ref{eq:19}), modulo the measurement protocol
\footnote{Eqn.(\ref{eq:22}), for the static and spherically symmetric
flow of (\ref{eq:35}), has circular orbits for which $({\bf v}_R(t).({\bf v}_R(t).{\bf  \nabla }).{\bf
v}({\bf r},t))=0$. There are also elliptical orbits, but because then $({\bf v}_R(t).({\bf v}_R(t).{\bf  \nabla
}).{\bf v}({\bf r},t))\neq 0$ this term
causes the ellipse to precess in the direction of motion (i.e. an advance of the perihelion),  and the rate of this 
precession is  determined in part by the eccentricity of  the orbit, but is zero for zero eccentricity.  However the
approximate analysis of (\ref{eq:33}) for elliptical orbits (see for example S. Weinberg, {\it Gravitation and Cosmology},
p.194) gives the well known form
$\Delta\phi=6\pi GM/c^2(1-e)a$ for the precession, but this is non-zero for circular orbits ($e=0$), which is a strange
result.  This suggests that there is a residual protocol or coordinate-time effect that arises in using the
Schwarzschild metric, and that the  analysis needs to be reviewed.  }.   Of course while the measurement protocol
has confused the nature of the dynamics encoded  in the GR equations, the GR equations nevertheless can be
helpful  when doing observations with electromagnetic signalling, since they then also encode the effects of the
quantum foam on the observers instruments, particularly time and/or frequency measurements, as for example in the
GPS navigation system.

\section{Conclusions}

Presented herein is the current state of development of  the radical proposal that to comprehend reality we
need a system richer than mere  syntax to capture the notion that reality is at all levels  about, what may be
called,  internal, relational or semantic information, and not, as in the case of syntax,
information that is essentially accessible to or characterisable by  observers, and that approach  amounts to the
assertion that reality can and should be modelled by symbol manipulating exercises, according to some externally
imposed set of {\it laws}.   This  necessitates an evolution in modelling reality from a non-process  physics to a
process physics.  Such a development has been long anticipated, with such insights dating back  $2500$ years 
to Heraclitus. 

 In particular process physics has now provided a derivation of Einstein's General Relativity model, as being a
manifestation of various effects of the emergent quantum foam system within process physics. The quantum foam is
the long sought for quantum theory of gravity, as encoded by the Quantum Homotopic Field Theory.  However the
higher level General Relativity, and its key  concept of {\it spacetime}, turns out to be essentially a phenomenological
construct, and not at all fundamental. The same applies to the technical aspect of General Covariance.  This
non-processing static   spacetime construct, with its representation of gravitational effects by means of
Riemannian curvature, turns out to be based on two essentially separate effects of the quantum foam, the first 
being the apparent invariance of the speed of light for any observer, which is caused by actual Lorenztian
dynamical effects upon clocks and rods, and which nevertheless provides the classical measurement protocol which
is  used in practical observations, but which is invalidated by non-local quantum processes of the Einstein,
Poldolsky and Rosen type.  The second effect is the effective diffusion/relaxation of the quantum foam towards
 quantum `matter', and this is the gravity effect. Amalgamating the two effects results in the spacetime
construct, but the curvature is induced by the peculiar aspects of the measurement protocol.  The genuine
gravitational effects may  be seen only after the removal of these effects.  In this way the Einstein field
equations then come down to assertions about the quantum foam velocity field behaviour.  Having exposed this
dynamics, and as well indicating the generic origin of the Lorentzian dynamical effects, the direction of future
research work in process physics becomes clear.   

To realise the process physics  one representation has been proposed and studied, namely that of a
self-referentially limited neural-network model, for neural networks are powerful examples of
non-symbolic information processing. This neural-network model is manifestly free of any notions
of geometry, quantum phenomena or even `laws of physics'.  Its origin  was  the observation that beneath
quantum field theory is essentially a neural network activity, and which now appears to be necessitated by the G\"{o}delian
limitations on syntactical systems, and the generalisation of theoretical physics modelling of reality from syntax to
semantics. The arguments presented here go a long way in demonstrating  that the phenomena of space, the quantum and the
time effects are emergent in the neural network, but only because the self-referential noise acts as a source of 
negentropy or order.  This self-referential noise is a new fundamental aspect of reality. It effects have  been apparent
ever since man became aware of  experiential time, with its dominant manifestation of the {\it Now}, and which physics has
so strongly denied.  Of course the enormous success of the geometrical modelling of time and its associated mechanical
clock-work determinism has long overwhelmed physics. Process physics has re-affirmed that time is
different to space. Poincar\'{e} and others, particularly the process philosophers, had noticed that determinism
implies, in principle, that since the `future' (or is it the `past'?) is entirely locked in, determinism forbids any
experiential {\it Now}. Einstein thought that this was unavoidable, but of course he was profoundly opposed to
intrinsic randomness, and so overlooked its enormous role in reality.  This  self-referential noise was also
apparent in the discovery of the randomness of the quantum measurement process, but the mechanical mindset of
physicists  was quickly reassured with the Born quantum measurement metarule; that we switch from wave phenomena to
Democritean `particles' to explain spots or clicks in detectors. 

The process physics model has been developed to the stage where various
phenomena have been identified and appropriate induced syntactical descriptions have been suggested. These correspond
essentially to the concepts of current physics which, over the years, have been arrived at via increasingly more abstract
non-process syntactical modelling.  One important addition being the ever-present QSD terms which, as it happens, ensure
that the phenomena of time fully matches our experiences of time, and which also plays various other key roles. Only by
abandoning the geometric model of time  has process physics been able to unify the phenomena of reality, and by
demonstrating that all the fundamental phenomena that interest physicists are necessarily emergent, that reality must be
just so, and not otherwise.  Indeed it is only by confronting the limits of logic and formalism do we actually arrive at an
understanding of how such modes of comprehending reality  have arisen and why they have been so effective, so much so that
their very effectiveness has blinded us to their limitations, and  the orthodoxy of the non-process physics approach had
become overbearing and even destructive to future physics. This new process-physics is inherently non-reductionist as it
explicitly assumes that reality is sufficiently complex that it is self-referential. 

Clearly we see the beginnings of a unification of physics that leads to quantum field theory, quantum
gravity and classicality and the emergence of syntax and its associated logic of named objects.  Such a
fundamental change in our comprehension of reality  will result in novel technological innovations, and one
of these, synthetic quantum systems and their possible role in `room temperature' quantum computers, has
been  suggested in   \cite{RC02}.

\end{document}